\documentclass[journal=jacsat,manuscript=article]{achemso}

\usepackage{chemformula} 
\usepackage[T1]{fontenc} 
\usepackage{siunitx}
\usepackage{amsmath}
\usepackage{amssymb} 
\usepackage{soul,xcolor}

\author{Johannes\,C. Rode}
\affiliation{Institut f\"ur Festk\"orperphysik, Leibniz Universit\"at Hannover, 30167 Hannover, Germany}
\altaffiliation{Contributed equally to this work}
\author{Dawei Zhai}
\affiliation{Departement of Physics \& Astronomy, Ohio University, Athens, OH 45701, USA}
\altaffiliation{Contributed equally to this work}
\author{Christopher Belke}
\author{Sung\,J. Hong}
\author{Hennrik Schmidt}
\affiliation{Institut f\"ur Festk\"orperphysik, Leibniz Universit\"at Hannover, 30167 Hannover, Germany}
\author{Nancy Sandler}
\affiliation{Departement of Physics \& Astronomy, Ohio University, Athens, OH 45701, USA}
\author{Rolf J. Haug}
\affiliation{Institut f\"ur Festk\"orperphysik, Leibniz Universit\"at Hannover, 30167 Hannover, Germany}
\email{haug@nano.uni-hannover.de}

\title
{Linking interlayer twist angle to geometrical parameters of self-assembled folded graphene structures}

\begin{document}

%
%
%
%
%

\begin{abstract}
Thin adhesive films can be removed from substrates, torn, and folded in distinct geometries under external driving forces. In two-dimensional materials, however, these processes can be self-driven as shown in previous studies on folded twisted bilayer graphene nanoribbons produced by spontaneous tearing and peeling from a substrate. Here, we use atomic force microscopy techniques to generate and characterize the geometrical structure of naturally self-grown folded nanoribbon structures.  Measurements of nanoribbon width and interlayer separation reveal similar twist-angle dependences possibly caused by the anisotropy in the bilayer potential. In addition, analysis of the data shows an unexpected correlation between the height of the folded arc edge -parameterized by a radius R-, and the ribbon width, suggestive of a self-growth process driven by a variable cross-sectional shape. These observations are well described by an energy minimization model that includes the bilayer adhesion energy density as represented by a distance dependent Morse potential. We obtain an analytical expression for the radius R versus the ribbon width that predicts a renormalized bending rigidity and stands in good agreement with experimental observations. The newly found relation between these geometrical parameters suggests a mechanism for tailored growth of folded twisted bilayer graphene- a platform for many intriguing physics phenomena.
\end{abstract}

\noindent \textbf{Keywords}: twisted bilayer graphene, graphene ribbons, graphene folds 
\vspace{.15 in}

External driving forces are needed to separate and fold thin films from substrates\cite{Hamm2008,Sen2010,Kruglova2011}, however, for the ultimate thin films, i.e. two-dimensional materials\cite{Novoselov2005,Geim2007}, these same processes  can be self-driven due to the growth of folded bilayer structures\cite{Annett2016}. The ultimate configuration is thus stabilized when the balance between the energies involved in bilayer formation, bending, peeling and tearing is reached. In general, the interlayer interaction between two-dimensional crystals is non-isotropic and depends on the interlayer lattice registry\cite{Berashevich2011A,Shibuta2011,Uchida2014}, factors shown to have a strong effect in self-driven structure configurations\cite{Wang2016,Zhu2017,Peymanirad2017}. At the same time, the different energies involved depend on the geometric factors of the structure (e.g. length, width, bending radius, etc). Therefore, stable configurations are expected to exhibit correlations between geometric parameters, that  may be used as a guide to specifically design bilayers and folded nano-arcs from self-assembly.

\begin{figure*}[htbp]
\includegraphics[width=6.4in]{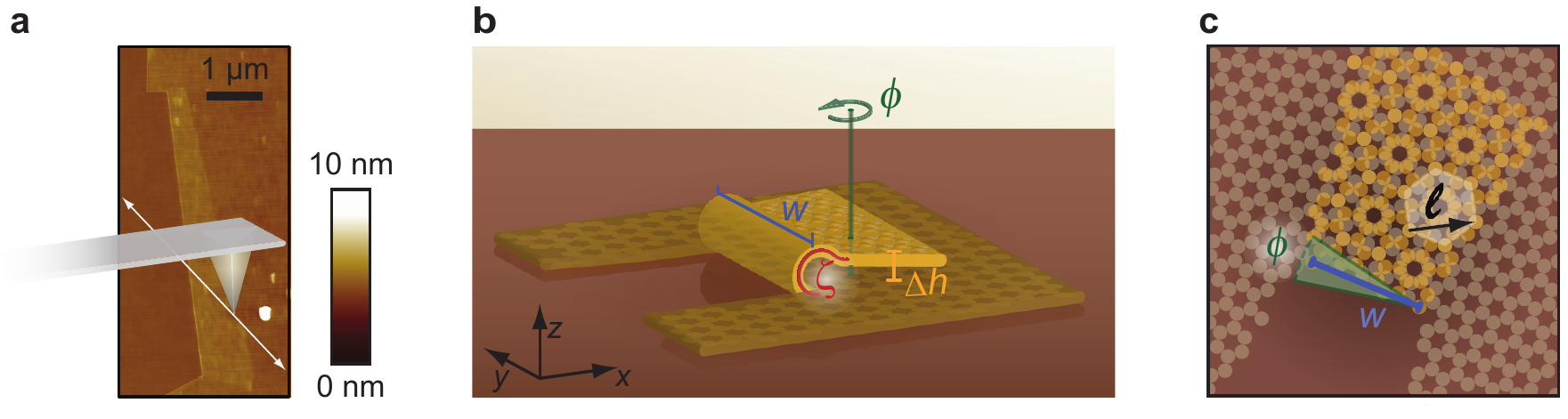}{\centering}
\caption{\textbf{Twisted bilayers of folded graphene ribbons.} \textbf{a} AFM topography of a fold in graphene; the stylized AFM probe and white arrow illustrate the process of nanomachining (AFM tip motion). Scale bars span \SI{1}{\micro\meter} in the lateral (black) and \SI{10}{\nano\meter} in the colored bar. \textbf{b}~Schematic of a ribbon folded out of graphene lying on a substrate with interlayer distance $\Delta h$ as well as width $w$, and cross-sectional arc $\zeta$ of the folded edge indicated in the figure. 
The interlayer twist angle $\phi$ between folded flap and mother flake is illustrated around the z-
axis of rotation.  \textbf{c} Atomic-scale schematic of folded graphene: in the TBG area, a commensurate superlattice unit cell with wavelength $\ell$ is highlighted as white hexagon; crystallographic axes (green) are mirrored at the folded edge and enclose the twist angle $\phi$.}
\end{figure*}

The central subject of our study are graphene ribbons folded out of monolayer graphene (MLG) prepared by mechanical exfoliation of natural graphite onto a silicon dioxide ($\mathrm{SiO_2}$) substrate.   
Folding events are initiated by scratching graphene via AFM nanomachining (see Methods), whereupon $\si{\micro\meter^2}$-sized bilayer areas emanate from the ruptured trench or the flake edge (Fig.\,1a). Ribbons generally tear out of the mother flake along two paths while staying connected to the bottom via a folded edge (Fig.\,1b). Results presented were obtained from the analysis of a set of 16 self-assembled ribbons, from 7 different flakes on top of 6 different substrates.
All experiments were carried out in ambient conditions.

Top and bottom lattices of these prepared structures will in general be rotationally misaligned in terms of their respective crystallographic directions (Fig.\,1c); {\it i.e.} ribbons and their substrate MLG constitute twisted bilayer graphene (TBG) structures.
This material system is host to a great variety of interesting physics\cite{Berashevich2011A,Shibuta2011,Uchida2014,Zhu2017,Peymanirad2017,Yamagishi2012,Kim2013,Schmidt2014,Rode2016,Rode2016arxiv,Rode2017,Jarrillo2018} originated in the geometric superposition of
the twisted lattices. Superstructures may be strictly periodic or incommensurate, depending on the corresponding rotational mismatch \mbox{angle $\phi$}\cite{Mele2010,dosSantos2012,Rode2017}.  

We have determined the interlayer twist angle $\phi$ (see Methods) for a number of self-assembled folded ribbons and observed that most values accumulate between \SI{20}{\degree} and \SI{30}{\degree}, as depicted in Fig.\,2a. 
Interestingly, the observed rotational mismatch coincides with low densities of commensurate interlayer configurations around \SI{21.8}{\degree} and \SI{27.8}{\degree} as marked in the corresponding parameter space shown in Fig.\,2b~\cite{Mele2010,dosSantos2012,Rode2017}.

The underlying reason is likely to be found in the self-assembled nature of this bilayer structure formation, as identified by Annett \textit{et al.}\cite{Annett2016}: 
starting from a folded-over flap with only nanometers of overlap onto the underlying MLG, the ribbon's growth is activated by thermal fluctuations and progressively stabilized by the gain in bilayer adhesion which overcompensates for the energy loss from tearing and peeling from the substrate.
Maximal expansion is typically reached on a timescale below the AFM image-frame acquisition. 

As the bilayer forms in a forward sliding motion, its growth process is favored by lower friction between the bottom and growing top layer. 
The extremely low friction condition, termed superlubricity\cite{Dienwiebel2004,Liu2012}, has been associated with incommensurate stacking configurations. 
Very low commensurate and fully incommensurate structures are most likely to be found in the range of values for $\phi$ observed predominantly in our self-assembled TBG (low density of small-wavelength commensuration, green backdrop in Fig.\,2b).
Alternatively, the assumption of an impeded growth of commensurate structures due to friction, is consistent with the absence of large numbers of commensurate bilayer graphene ribbons (Figs.\,2a,b).

\begin{figure}[htpb]
\includegraphics{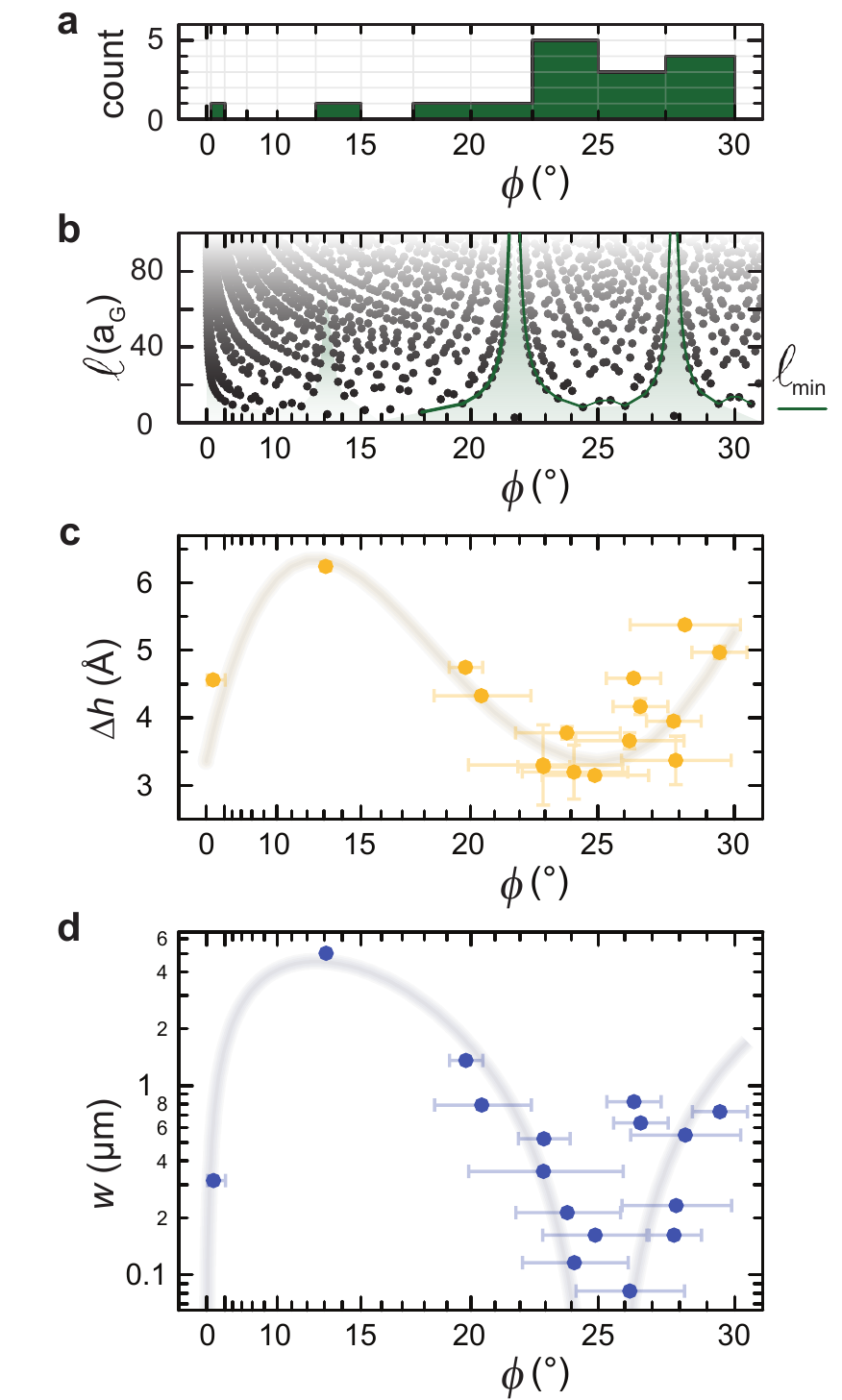}{\centering}
\caption{\textbf{Rotational mismatch in folded bilayer ribbons.} \textbf{a} Histogram counting the number of folded ribbons per increment of rotational mismatch angle $\phi$. The horizontal axis  is scaled with $\cos(3\cdot\phi)$ in order to increase discernibility in areas of accumulation around larger interlayer twists. \textbf{b} Commensurate configurations in terms of wavelength $\ell$ (in units of graphene's lattice constant $a_\mathrm{G}=\SI{2.46}{\angstrom}$) vs. $\phi$. The green shading highlights areas of low density of commensurate structures. The contour of $\ell_\mathrm{min}$ (green line) connects to the smallest $\ell$ within increments of \SI{1}{\degree} or less, disregarding the isolated small-wavelength configurations at \SI{21.8}{\degree} and \SI{27.8}{\degree}. \textbf{c} Interlayer distance $\Delta h$ vs. $\phi$. The gray line serves as a guide to the eye. \textbf{d} Width of the folded edge $w$ vs. $\phi$. The gray line is a sinusoidal fit serving as a guide to the eye.}
\end{figure}

Rotational mismatch does not only determine the growth probability of a self-assembled ribbon but is also a quantitative predictor of its final geometrical structure: 
Figure~2c depicts the samples' interlayer distance $\Delta h$ which is extracted as the difference between ribbon and MLG heights from AFM topography measurements (see Methods). The dependence on the interlayer twist displays an oscillating behavior (gray line, Fig.\,2c) with a pronounced dip around $\phi=\SI{25}{\degree}$. Note that measured values for $\Delta h$ lie between $\SI{3.2}{\angstrom}$ and $\SI{6.2}{\angstrom}$, around a median of $\SI{4.1}{\angstrom}$, beyond theoretically anticipated variations for naturally occurring bilayer structures\cite{Berashevich2011A,Shibuta2011,Uchida2014,Peymanirad2017}. Examples of similarly large interlayer distances have been found in artificially stacked graphene  in previous experimental works: 
values for $\Delta h$ from \SIrange{3.4}{4.1}{\angstrom}\cite{Yamagishi2012}, and around \SI{4}{\angstrom}\cite{Geim2007,Kim2013} have been reported for TBG produced from transfered and folded samples respectively. 

While it is plausible that the discrepancy between theory and experiment lie in a sparse ordered intercalation of adatoms keeping top and bottom layers apart at a uniform distance, a more intrinsic mechanism involving superlubricity appears more likely. We propose that
the existence of a superlubric state\cite{Dienwiebel2004,Liu2012}, facilitated by lack of interlayer commensuration, enables the growth of these ribbons in the first place\cite{Annett2016}.
Unfortunately, there is a lack of appropriate theory models to describe these incommensurate stacking structures because of the absence of periodicity, in stark contrast with lattice-periodic structures\cite{Uchida2014}\cite{Berashevich2011A,Shibuta2011}.
Notice that, whichever reason applies, i.e. intercalation or superlubricity, the effect correlates with interlayer twist, suggesting a causal connection between the minimum in $\Delta h$ and the dip in $\ell_\mathrm{min}$, situated around a value of relative interlayer orientation of $\phi=\SI{25}{\degree}$ (Figs.\,2b,c).

We have observed another interesting correlation between the orientational mismatch and the width of a folded nanoribbon in its final configuration,
defined in Fig.\,1b as
$w$.  With 
values that 
vary between $\SI{0.08}{\micro\meter}$ and $\SI{5.01}{\micro\meter}$, $w$ oscillates as a function of the interlayer twist, and displays a pronounced decline towards a minimum value of $\phi=\SI{25}{\degree}$ (Fig.\,2d), similar to the dependence observed for $\Delta h(\phi)$.

To understand the origin of these dependencies, we turn to the internal energy $U$ of the torn and folded system (see Supplementary Information):
from the vanishing of the partial derivative $\partial U/\partial L_\mathrm{gg}$, where $L_\mathrm{gg}$ is the length of the growing bilayer graphene area (see Fig.\,4c), one finds          
\begin{equation}
\label{eqWvsDE}
w=\frac{2\lambda}{\gamma_{gg}-\gamma_{gs}},
\end{equation}
where $\gamma_{gg}$, $\gamma_{gs}$ and $\lambda$ are the graphene-to-graphene adhesion energy density, graphene-to-substrate adhesion energy density, and tear energy density per ruptured path, respectively. Notice that dissipative contributions due to friction are neglected in the derivation of the expression above, as we assume a superlubric TBG interaction. This relation corresponds to the one obtained in Ref.~\citenum{Annett2016} for taper angles $\theta \simeq 0$, consistent with measured values 
$\cos{\theta}=(\num{0.98}\pm\num{0.01})\approx1$ across the whole range of our samples.


According to Eq.~\ref{eqWvsDE}, the twist-angle dependence of $w$ may be caused by either of the contributing energy densities, evidence of the high sensitivity of TBG superstructure formation to the specific values of twist angles\cite{Mele2010,dosSantos2012,Rode2017} (see Fig\,2b). However, a closer analysis suggests the bilayer adhesion energy density $\gamma_{gg}$ to be the most probable cause for the dependence observed. To a lesser degree, $\lambda$ is expected to vary slightly, depending on the direction of tearing paths consistent with the measured values of taper angles, while the adhesion energy density $\gamma_{gs}$ between graphene and the amorphous substrate is largely isotropic. 

In principle, the bilayer interaction depends on both the twist angle and the interlayer separation, however as Fig.\,2c shows, these are not independent variables for self-grown structures making $\gamma_{gg} = \gamma_{gg}[\Delta h(\phi)]$. 
The relation between $w$ and $\Delta h$ as obtained from the measured samples is analyzed in Fig.\,3 that shows data accumulating around the diagonal of the double-logarithmic plot window. 
This correlation is consistent with the similar $\phi$-dependencies measured for both magnitudes shown in Fig.\,2c, 2d. 
Note that a local minimum in $w$ may be identified around $\Delta h=\SI{3.5}{\angstrom}$.

\begin{figure}[hbtp]
\includegraphics{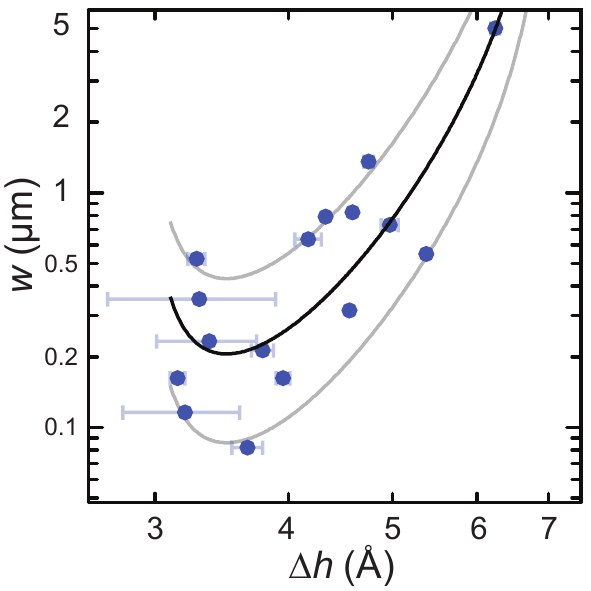}{\centering}
\caption{\textbf{Connecting \textit{w} and $\Delta h$ via the Morse potential.} 
Width of the folded edge \textit{w} vs. interlayer distance $\Delta h$. The black line corresponds to
a fit after Eq.~\ref{eqWvsDE} (substituting $\gamma_\mathrm{gg}$ from Eq.~\ref{eqMorse}). Parameters of the Morse potential are
fitted as $M=\SI{0.215}{\joule\meter^{-2}}$, $\sigma=\SI{1.25}{\angstrom^{-1}}$ and $\Delta h_0=\SI{3.50}{\angstrom}$, while energy densities 
have values of $\gamma_\mathrm{gs}=\SI{0.005}{\joule\meter^{-2}}$ and $\lambda=\SI{21.5}{\nano\joule\per\meter}$ with the constraints described in the Supplementary Information. Gray lines correspond to different values of tearing energy densities of $\lambda=\SI{9}{\nano\joule\per\meter}$ (bottom) and $\lambda=\SI{45}{\nano\joule\per\meter}$ (top) for the same values of other parameters.}
\end{figure}

To quantify these observations we model the bilayer interaction energy density $\gamma_{gg}$ with the interlayer-distance dependent Morse potential
\begin{equation}
\label{eqMorse}
\gamma_\mathrm{gg}(\Delta h)=-M\cdot\left\lbrace\left[1-e^{-\sigma\cdot(\Delta h-\Delta h_0)}\right]^2-1\right\rbrace,
\end{equation}
where $M$ defines the maximum adhesive potential strength, $\sigma$ adjusts for the spatial extent of the potential and $\Delta h_0$ defines the optimal layer separation. We use this expression to represent $\gamma_\mathrm{gg}$ in Eq.~\ref{eqWvsDE}, and obtain the function $w(\Delta h)$ to fit the data. The quantities $M$, $\sigma$ and $\Delta h_0$, treated as free parameters,  were constrained by corresponding fits of Eq.~\ref{eqMorse} to theoretically predicted interlayer potentials for AB- and AA-stacked graphene bilayers respectively\cite{Peymanirad2017} (see Supplementary Information).

The resulting fitting function is displayed as the black curve in Fig.\,3: starting at small interlayer distance, the folded width declines up to a minimum, corresponding to a maximal value for $\gamma_\mathrm{gg}$ beyond the dominion of repulsive short-range interaction. Increasing values of $w$ for $\Delta h>\SI{3.5}{\angstrom}$ reflect the waning of the attractive long-range contributions due to an increased interlayer separation.  

The corresponding fitting parameters are listed in the caption of Fig.\,3: the adjusted values of the Morse potential obtained in the numerical procedure lie in the middle of the range corresponding to those calculated for AA- and AB-stackings\cite{Peymanirad2017} (see Supplementary Information).  
Thus, they are consistent with theoretical predictions for less ordered stacking configurations\cite{Peymanirad2017,Shibuta2011,Uchida2014} as those expected in folded ribbons\cite{Annett2016}. 
The adjusted adhesion energy density $\gamma_\mathrm{gs}$ of graphene on $\mathrm{SiO_2}$ lies at the lower end of reported values\cite{Kusminskiy2011,Ishigami2007,Koenig2011} and fits theoretical expectations for an interlayer of water between sample and substrate\cite{Kusminskiy2011}, expected in our ambient-conditions setup. 
Finally, the adjusted value for tearing energy density, $\lambda=\SI{2.15e-8}{\joule\per\meter}$, lies somewhat above the expected theoretical minimum of \SI{2.8e-9}{\joule\per\meter} (obtained assuming a straight cut along the zigzag direction at one C-C bond of \SI{4.3}{\electronvolt} per lattice constant $a_\mathrm{G}=\SI{2.46}{\angstrom}$). Possible explanations for this discrepancy may lie in the nanoscale structure of the tearing length where 
torn paths, seemingly straight within the resolution of the AFM, could meander back and forth, increasing the edge length; 
or several bonds may tear in parallel, e.g. in response to spontaneous strain release, thereby widening the tearing path. 
Support for these scenarios is provided by the variation in $\lambda$ from sample to sample: when allowing for a certain range of values for $\lambda$, as mentioned in the caption of Fig.\,3, the full scattering in the data is accounted for while maintaining an unchanged set of the remaining fitting parameters (gray lines at the bottom and top of Fig.\,3). 


\begin{figure*}[htpb]
\includegraphics[width=6.4in]{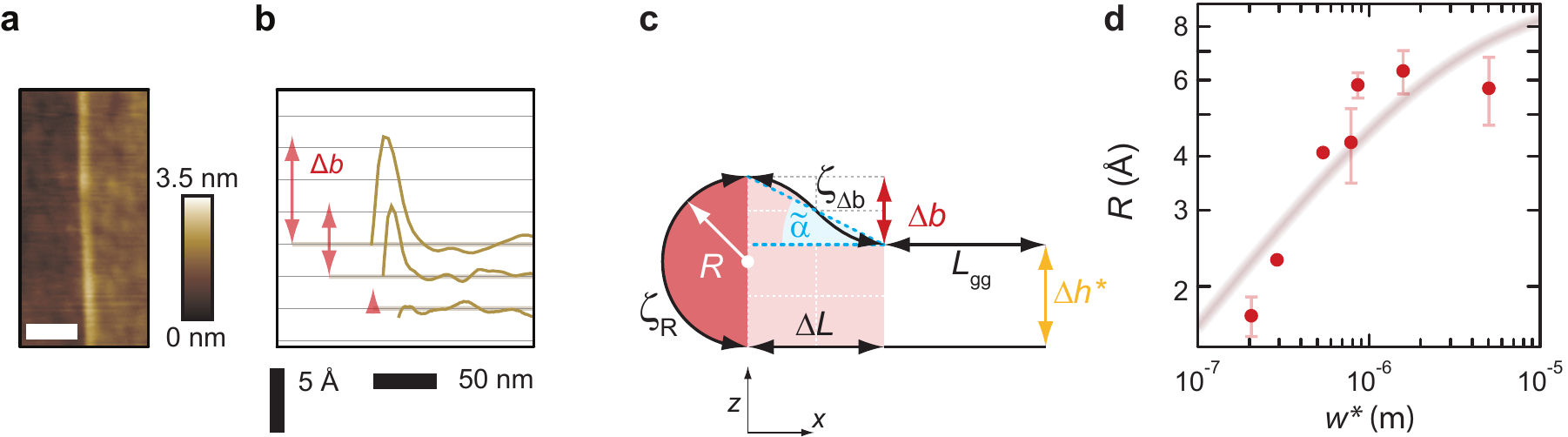}{\centering}
\caption{\textbf{Cross-sectional profile of the folded edge.} \textbf{a} AFM topography of substrate (left), TBG plane (right) and protruding folded edge (middle). The scale bar (white) spans a distance of 100\,nm. \textbf{b} Cross-sections of height profiles measured across three folded edges of different bump heights $\Delta b$. \textbf{c} Schematic cross-section of a folded structure with adhered bilayer length $L_\mathrm{gg}$ and folded edge with partitions of the bent arc $\zeta_\mathrm{\Delta b}$ and $\zeta_\mathrm{R}$ as well as flat part $\Delta L$. \textbf{d} Bending radius $R$ as calculated from measured $\Delta b$ and $\Delta h^*$ vs. width of the folded edge $w^*$; the curved line is a fit after eq.~\ref{eqR}.}
\end{figure*}

Finally, the bended edge connecting top and bottom layer is unique to the folding approach in the formation of the TBG. 
In terms of electronics, it is e.g. predicted to give rise to snake states\cite{Rainis2011} and has been linked to transport features independent of perpendicular magnetic field\cite{Schmidt2014}.

Due to a finite bending rigidity $D$, the folded graphene arc usually protrudes above the TBG plane as resolved in AFM topography (Fig.\,4a). 
We find the bump height $\Delta b$ to vary between virtually zero and up to \SI{8.5}{\angstrom} (Fig.\,4b). 
To relate the folded section to the above discussed structural parameters, the full arc $\zeta$ is divided in two contributions $\zeta_\mathrm{\Delta b}$ and $\zeta_\mathrm{R}$ as depicted in Fig.\,4c. The quantities $\Delta b$ and bending radius $R$, are related to the interlayer distance $\Delta h$ via   
\begin{equation}
\label{eq:2}
R=(\Delta h+\Delta b)/2.
\end{equation}  
From this expression we deduce magnitudes for $R$ between \SI{1.7}{\angstrom} and \SI{6.3}{\angstrom}. {\it These values exhibit a strong correlation with the measured values for the corresponding ribbon's width $w$, revealing that the self-growth process is driven by a variable cross-sectional shape}. The newly found relation corresponds to a different self-growth regime from the one identified
in previous experimental studies\cite{Annett2016} 
where self-directed growth was dominated by changing widths and assumed to occur with constant cross-sectional shapes (fixed binding energy), or 
with 
focus 
on the bending stiffness\cite{Chen2014,Chen2015}. On the other side, theory studies have addressed conditions for fold formation such as growth beyond a minimal critical length independent of ribbon width\cite{Meng2013}, or fixed interlayer separation of self-folded mono and multi-layer graphene\cite{Cranford2009}, 
with an analysis of transport consequences of folded arcs\cite{Rainis2011}. We note that neither of these are applicable to the ribbons reported in this work.

As a consequence, a different picture emerges when a full edge profile of a self-grown ribbon is considered. In this approach, $\Delta h$ and $w$ (related via Eqs.\,1 and 2) are considered fixed quantities $\Delta h^*$ and $w^*$ within a given sample, both determined by the interlayer twist angle (Fig.\,2c, 2d). 
The folded arc is modeled as a semi-circle of length $\zeta_\mathrm{R}=\pi R$, where the descending part of length $\zeta_\mathrm{\Delta b}$ is, for simplicity, parametrized by two independent variables ${\Delta b=2R-\Delta h^*}$ and $\Delta L$ (see Fig.\,4c). 
Note that the condition imposed by Eq.\,1 remains unaffected by these choices (see Supplementary Information for further details). 
After using Eq.\,1 to simplify the expression for the internal energy $U$, we require $\partial U/\partial R=0$ in the equilibrium condition,  obtaining the relation:   
\begin{equation}
\label{eqdUdR}
\gamma_\mathrm{gg}\left(\pi+\frac{\partial\zeta_\mathrm{\Delta b}}{\partial R}\right)-\frac{D}{2R^2}\left(\pi-R^2\frac{\partial f}{\partial R}\right)=0,
\end{equation}
where $f(\Delta b,\Delta L)$ represents the path integral along $\zeta_\mathrm{\Delta b}$ over the curved region, determining the bending energy associated with the descending arc $U_{\Delta b}=\frac{Dw}{2}f(\Delta b,\Delta L)$. 
This contribution can be parameterized by an effective curvature $\tilde{\kappa}_{\Delta b}$ that can be compared with the one determined by the fixed radius $R$. 
It is useful also to introduce an effective angle $\tilde{\alpha}$ defined by $\Delta b \approx \zeta_\mathrm{\Delta b} \sin\tilde{\alpha}$ that, combined with Eq.~\ref{eq:2}, renders $\partial\zeta_\mathrm{\Delta b}/\partial R \approx 2/\sin\tilde{\alpha}$ (see Fig.\,4c). 
In the regime $\tilde{\kappa}_{\Delta b}\ll R^{-1}$, we find
\begin{equation}
\label{eqR}
R\approx\sqrt{\frac{\pi D}{2\gamma_\mathrm{gg}\left(\pi+2/ \sin\tilde{\alpha}\right)}}=\sqrt{\frac{\tilde{D}w^*}{4\lambda+2\gamma_\mathrm{gs}w^*}},
\end{equation}
where $\tilde{D}=\frac{\sin\tilde{\alpha}}{\sin\tilde{\alpha}+2/\pi}D$ is the renormalized bending rigidity coefficient (see Supplementary Information).

The dependency of $R$ on $w^*$ predicted above, is confirmed in our data as shown in Fig.\,4d:  a fit of Eq.\,5 (gray line) captures fairly well the rising trend in bending radius with increasing folded width, including the onset of saturation for large values of $w^*$ (the data presented correspond\st{s} to a subset of seven ribbons only, as accurate measurement of the narrow folded arc requires special care in AFM operation). 

A key to the qualitative understanding of this behavior is the competition between contributions from bending, adhesion, and tearing energies to the equilibrium condition $\partial U/\partial R=0$. 
Whereas the minimization of bending energy is achieved towards larger values of $R$, minimization of the tearing as well as adhesion energies require a decreasing bending radius. In narrow ribbons (large edge-to-bulk ratio), the equilibrium condition is dominated by the contribution from the tearing edges, which is reflected in small optimal bending radii at small values of $w^*$.
In wide ribbons, the equilibrium condition is dominated by the counteracting contributions from bending and adhesion alone, as both corresponding energies scale proportionally with $w^*$ (in contrast to tearing energy); this renders values of $R$ effectively independent of folded width only at large values of $w^*$. 

Quantitatively, the renormalization of the bending rigidity coefficient $D$, accounting for the contributions from the descending part of length $\zeta_\mathrm{\Delta b}$, plays a crucial role in understanding the magnitude of values of $R$: From the fitting in Fig.\,4d, we find $\tilde{D}=\SI{0.12}{\electronvolt}$, for values of $\lambda=\SI{18.3}{\nano\joule\per\meter}$ and $\gamma_\mathrm{gs}=\SI{0.011}{\joule\meter^{-2}}$ obtained with the same fitting procedure used in Fig.\,3, but applied on the subset of seven ribbons shown (see Supplementary Information). 
A comparison with the reported value for $D\approx\SI{1}{\electronvolt}$\cite{Wei2013}, renders typical values of $\tilde{\alpha}\approx\SI{5}{\degree}$. 
This implies extensions of the order of $\zeta_\mathrm{\Delta b} \lesssim\SI{10}{\nano\meter}$ for the descending arc (Fig.\,4c), similar in magnitude to the typical tip radius of the AFM probes used.
The apparent lateral extensions of the bumps shown in Fig.\,4a, 4b are thus consistent (within the corresponding lateral resolution) with the proposed model and calculated values.    

\section{Conclusion}
In conclusion, our study points toward a pivotal role of interlayer commensuration in twisted bilayer morphology. 
The strong dependence of the interlayer adhesion energy density with twist angles and corresponding interlayer separations reveals a rich parameter space that can potentially be explored during the growth of folded graphene ribbons. 
Refinement in nanoindentation techniques in conjunction with a more precise knowledge of sample crystallography may allow for control over interlayer twists. 
This may ultimately enable tailored ribbon synthesis at preplanned layer separation, as well as customized folded arcs which could be exploited e.g. in devices with built-in ultrahigh pseudo-magnetic fields.
Moreover, folded sandwich-structures including an insulating layer (e.g. hexagonal boron nitride) will enable three-dimensional device structures with defined current flow along the cross-sectional arc of the folded edge, and have potential application in the developement of rolled-up capacitors.

\section{Methods}

\textbf{AFM nanomachining.}
Nuclei for fold-growth are seeded by scratching a graphene flake via AFM nanomachining, thereby creating additional edge-surface with rough borders which are prone to fold-overs (Figure\,1a).
Reliable results are achieved by repetitive contact-mode tracing (some ten repetitions) with a high spring-constant, diamond-coated probe, operating in the \si{\micro N}-range of spring load.

\textbf{Twist-angle determination.}
In TBG prepared by folding of a monolayer, the angle $\phi$ between different crystalline directions in the top and bottom layers can, in general, be deduced from sample geometry\cite{Schmidt2014,Rode2016arxiv,Rode2017}: 
the folded edge (blue line in Figure\,1c) acts as a mirror axis between crystalline symmetry directions (green lines in Figure\,1c) in the bottom and folded-over layer respectively.
As graphene flakes are terminated by straight armchair- or zigzag-edges in the majority of cases\cite{Geim2007,Neubeck2010}, a set of clean facets in $n\times\SI{30}{\degree}$-increments ($n$ being an integer) are indicative of the crystallography in a given sample. 
This allows to obtain the magnitude of $\phi$, via doubling of the angle between the folded edge and the straight flake facet.
Depending on image resolution and length of the edges, the accuracy of this geometric approach is $\pm\SI{1.5}{\degree}$.
Combination of the honeycomb lattice's \SI{120}{\degree} rotational symmetry, and  mirror symmetries about the two crystalline directions -armchair and zigzag- renders an  angle $\phi$ within the range  $\phi\in[\SI{0}{\degree},\SI{30}{\degree}]$. As a consequence, structures at $\phi$ and \SI{60}{\degree}-$\phi$ are herein identical except
for a possible translational shift between top and bottom lattice, depending on the axis of rotation\cite{Mele2010,dosSantos2012,Rode2017}. The determination of the twist angle via AFM-images is given in the SI-Figures 5-8 for some selected samples.

\textbf{Extraction of interlayer distance.}
To extract the vertical separation $\Delta h$ between 
top and bottom layers of the planar TBG section, we record AFM topography over the folded ribbon and surrounding MLG (Supplementary Fig.\,1a). Measurements are undertaken in contact mode, special care is taken to minimize artifacts from mechanical crosstalk that may arise in step measurements over surfaces of different frictional coefficients\cite{Warmack1994,Hoffmann2007}: triangular Silicon Nitride probes of $\sim\SI{0.3}{\per\newton}$ force constant are employed for torsional and buckling stability, the scanning speed is limited to \SI{2}{\micro\meter\per\second}, and remaining differences between topography information from trace and retrace directions are averaged out by adding both channels' data and dividing by two.
The topography information is evaluated via a histogram that counts the number of pixels per interval of height $h$ as shown in Supplementary Fig.\,1b (gray dots). Depicted with a black line, a sum of two Gaussian distributions,
\begin{equation}
f(h)=\sum_\mathrm{i\in\{MLG,TBG\}}C_\mathrm{i}\cdot\exp{\left(-\frac{h-h_\mathrm{i}}{\sigma_\mathrm{i}}\right)^2},
\end{equation}                 
is used to fit the data. Individual contributions from MLG and TBG are plotted in brown and orange respectively. Interlayer distance is extracted from fitting results as $\Delta h=h_\mathrm{TBG}-h_\mathrm{MLG}$, where the error is defined as sum of fitting uncertainties in $h_\mathrm{TBG}$ and $h_\mathrm{MLG}$.

\begin{acknowledgement}

The authors acknowledge financial support from the DFG within the priority program SPP 1459, the School for Contacts in Nanosystems, and the "Fundamentals of Physics and Metrology" initiative (JCR, CB, SJH, HS, and RJH), NSF-DMR 1508325 (DZ and NS). This work was partially performed at the Aspen Center for Physics, which is supported by NSF grant  PHY-1066293 (NS). J. C. Rode acknowledges support from the Hannover School for Nanotechnology. The authors thank Peter Behrens and Hadar Steinberg for helpful discussion.

\end{acknowledgement}

%
%

\bibliography{Maintxt_Refs}

\end{document}